\documentclass[aps,twocolumn,showpacs,english]{revtex4-1}
\usepackage{lmodern}
\usepackage[T1]{fontenc}
\usepackage[latin9]{inputenc}
\usepackage{textcomp}
\usepackage{amsmath}
\usepackage{graphicx}
\usepackage{SIunits}
\usepackage{color}

\usepackage{babel}
\usepackage[unicode=true,pdfusetitle,
 bookmarks=true,bookmarksnumbered=false,bookmarksopen=false,
 breaklinks=true,pdfborder={0 0 1},backref=false,colorlinks=false]
 {hyperref}
\begin{document}
\preprint{APS/123-QED}

\title{Giant non-linear
interaction between two optical beams via a quantum dot embedded in a photonic wire}

\author{H.A. Nguyen$^{1,2}$, T. Grange$^{1,2}$, B. Reznychenko$^{1,2}$, I. Yeo$^{1,2,3}$, P.-L. de Assis $^{4}$, D. Tumanov$^{1,2}$, F. Fratini$^{1,2}$, N.S. Malik$^{1,3}$, E. Dupuy$^{1,3}$, N. Gregersen$^{5}$, A. Auff\`{e}ves$^{1,2}$,
  J.-M. G\'{e}rard$^{1,3}$, J. Claudon$^{1,3}$,  and J.-Ph. Poizat$^{1,2,*}$}

\affiliation{$^1$ Univ. Grenoble Alpes, CNRS, Grenoble INP, Institut NEEL, F-38000 Grenoble, France \\
$^2$ Univ. Grenoble Alpes, CNRS, Grenoble INP, Institut NEEL, "Nanophysique et semiconducteurs" group, 38000 Grenoble, France \\
$^3$ Univ. Grenoble Alpes, CEA, INAC, PHELIQS, "Nanophysique et semiconducteurs" group, F-38000 Grenoble, France \\
$^4$ Gleb Wataghin Institute of Physics,  University of Campinas - UNICAMP, 13083-859, Campinas, S\~{a}o Paulo, Brazil \\
$^5$ Department of Photonics Engineering, DTU Fotonik, Kongens Lyngby, Denmark \\
* Corresponding author :  jean-philippe.poizat@neel.cnrs.fr}

\date{\today}% It is always \today, today,

\begin{abstract}
Optical non-linearities usually appear for large intensities, but discrete transitions allow for giant non-linearities operating at the single photon level. This has been demonstrated in the last decade for a single optical mode with
 cold atomic gases, or  \emph{single} two-level systems coupled to light via a tailored photonic environment.
  Here we  demonstrate a \emph{two-modes} giant non-linearity by using a three-level structure in a single
  semiconductor quantum dot (QD) embedded in a photonic wire antenna.
  The large coupling efficiency and the broad operation bandwidth of the photonic wire enable us
  to have two different laser beams interacting with the QD in order  to control  the reflectivity of a laser beam with the other one using as few as $10$ photons per QD lifetime. We discuss the possibilities offered by this easily integrable system for ultra-low power logical gates and optical quantum gates.
\end{abstract}

\maketitle

Whether classical or quantum, optical communication has proven to be the best approach for long distance information distribution. All-optical data processing has therefore raised much interest in recent years, as it would avoid energy and coherence consuming optics-to-electronics conversion steps \cite{Miller10}.
Optical logic requires  non-linearities  to enable photon-photon interactions \cite{Turchette95PRL,Chang07,Reiserer,Tiecke14,Sun16,Maser16,Hacker16,Shomroni17}.
%Two-ports operation is a necessary requirement for the implementation of any non-trivial optical data processing. This involves a non-linear interaction between two distinct optical modes.
Low power optical logic faces the challenge of implementing  non-linear effects that usually occur at high power.
Interestingly, such  functionalities ultimately operating  at the single photon level can be achieved with  giant non-linearities  obtained via resonant interactions
%with optically well-coupled atomic-like
with system featuring discrete energy levels  \cite{Turchette95PRL,Alexia07,Chang07,Maser16,Hacker16,Shomroni17,Birnbaum05,Englund07,Fushman08,Hwang09,Aoki09,Rakher09,Loo12,Volz12,Englund12,Bose12,Lukin13,Rempe14PRL,Chang,Lodahl,Javadi15,Feizpour15,deSantis17,Reiserer,Tiecke14,Sun16}.
Apart from refs \cite{Lukin13,Feizpour15}  that exploit electromagnetic induced transparency of an atomic cloud, most of the experimental realizations use the concept of  "one-dimensional atom" \cite{Turchette95APB}, wherein a single two-level system is
predominantly coupled to a single propagating spatial mode.

 During the last decade, "one-dimensional atoms" have been implemented  with  single atoms \cite{Turchette95PRL,Birnbaum05,Aoki09,Tiecke14,Rempe14PRL,Reiserer,Hacker16,Shomroni17},  molecules \cite{Hwang09,Maser16} or  semiconductor QDs \cite{Englund07,Rakher09,Volz12,Loo12,Englund12,Bose12,Sun16,Fushman08,Chang,Lodahl,Javadi15,deSantis17}
 thanks to  resonant optical cavities \cite{Birnbaum05,Englund07,Aoki09,Rakher09,Volz12,Loo12,Englund12,Bose12,Tiecke14,Reiserer,Lodahl,Sun16,Hacker16,Shomroni17,deSantis17},  as well as  via the interaction with a tightly focused optical beam  \cite{Hwang09,Maser16} or non-resonant field confinement such as in single transverse mode waveguides \cite{Chang07,Javadi15},
Whereas \emph{single} mode giant non-linearity has been demonstrated with various systems, optical computing requires the non-linear interaction between \emph{two} different optical channels \cite{Maser16,Englund12,Bose12,Hacker16,Shomroni17}.
%Recent results have shown  that  \emph{two-modes} interactions can be mediated by a single atom or QD coupled to an optical cavity \cite{Englund12,Bose12,Hacker16,Shomroni17} or by a single molecule at the waist of a tightly focused beam \cite{Maser16}.
%a two modes cross non-linearity  can be implemented with a three-level structure wherein few photons in one mode control the reflection/transmission of light in the other one \cite{Hoi11,Volz12,Englund12,Bose12,Maser16}.

\begin{figure}%[t]
\includegraphics[width=\linewidth]{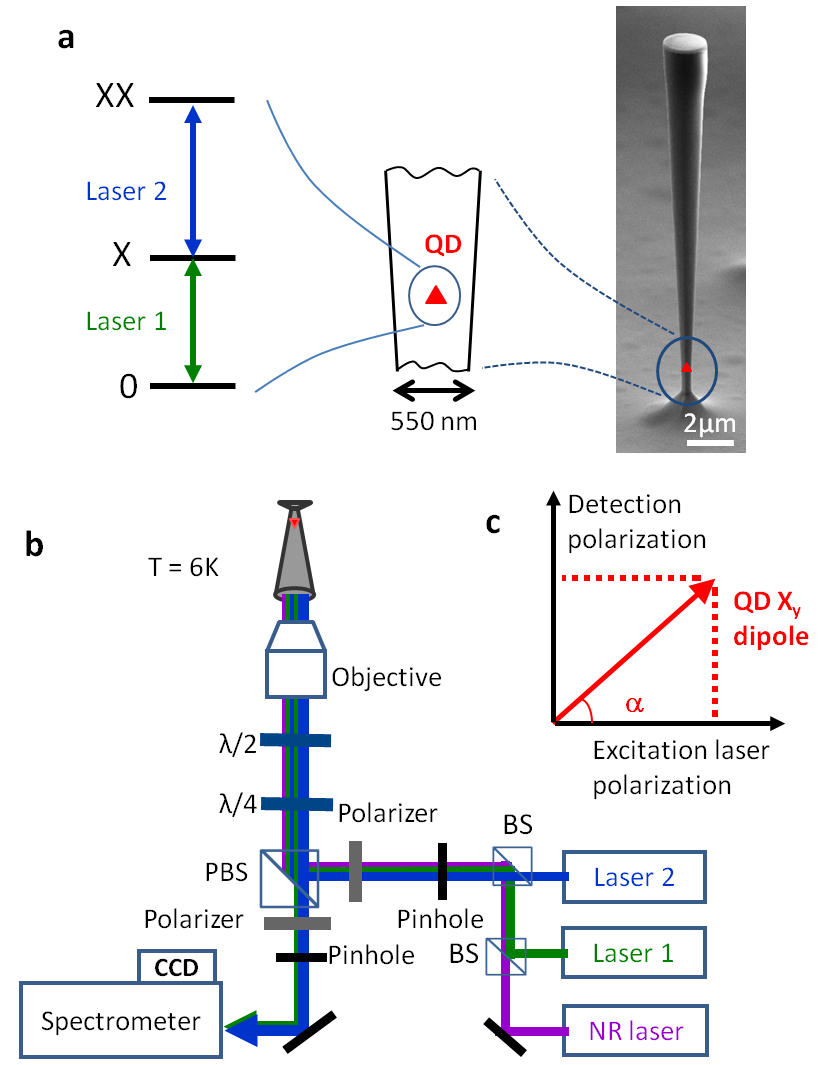}
\caption{\textbf{Sample and set-up.} \textbf{a}, On the right is shown a scanning electron microscope image of the sample. The central part details the location of the QD at the bottom of the wire where the diameter provides an efficient coupling of the QD spontaneous emission to the fundamental guided mode.
% offers a large  $\beta$ ratio of the spontaneous emission rate in the waveguide normalized to the total spontaneous emission rate.
The empty QD, the excitonic (X) and biexcitonic (XX) states form a non-degenerate three-level ladder scheme as represented on the left.  The two transitions are addressed by two different lasers.
\textbf{b}, Experimental set-up.
Two  tunable continuous wave  lasers excite the QD transitions.     The lasers are spatially filtered with a pinhole located at the focal point of a lens. They are then focused on the sample with a microscope objective of $0.4$  numerical aperture. A confocal detection with a pinhole   selects only the light coming out of the photonic wire. An extra non-resonant (NR) laser is used for photoluminescence spectroscopy, and as a "quietening  laser" during the experiments \cite{SM}.
A polarizing beam splitter (PBS) is used for a cross-polarized detection scheme.
\textbf{c}, The lasers are linearly polarized at an angle $\alpha=27^o$ with respect to the QD dipole of interest. The detection is performed along the polarization orthogonal to the laser polarization. Waveplates  in front of the objective ensure a precise control of the lasers polarization \cite{Kuhlmann13}. }
\label{fig:1}
\end{figure}

In this letter, we demonstrate  a \emph{two-modes} cross non-linearity in an alternative semiconductor system that offers the perspective of being
perfectly compatible with photonic circuits based on planar waveguides, and therefore easily integrable.
 We use the  three level structure of the biexciton-exciton (XX-X) scheme \cite{Moreau} in a semiconductor   QD embedded in a  waveguide antenna \cite{JClaudon,Kolchin11,Munsch13,Hayrynen16} (see Fig. \ref{fig:1}a and Supplemental Material \cite{SM}).
%A similar device has demonstrated record efficiency as a single photon source \cite{Munsch13}.
Thanks to the broadband nature of the waveguide, as opposed to microcavities, both excitonic and biexcitonic transitions of a QD located at the center of this structure are efficiently coupled to its fundamental guided mode.
We show that turning on the control beam with a power as small as 1 nW alters significantly the reflection of  the probe beam at a different wavelength. We present
two different schemes depending on whether the control (probe) beam is tuned around the lower (upper) or the upper (lower) transition. We expose their different physical mechanisms and performances and discuss their respective potentials for ultra-low power optical computation and for photonic quantum computation.

% Ce paragraphe va dans la descrition du système expérimental

%A QD can host several excitons. The two transitions involved in the three-level ladder structure used in this experiment are the biexcitonic and excitonic transitions (see Fig.XX)

 The investigated structure (Fig. \ref{fig:1}a) is  a vertical GaAs photonic wire containing an InAs QD. The QD is located at the bottom of the wire where its diameter ($500$ nm) ensures a good QD coupling to the fundamental guided mode \cite{JClaudon,Hayrynen16}. This mode expands as it propagates along the conical taper before being out-coupled \cite{Munsch13}. The photons of both transitions can then be efficiently collected by a microscope objective and their spatial mode features a large overlap with  a Gaussian beam \cite{Stepanov_APL}. For the same reasons,  focusing  a mode-matched Gaussian beam on the top facet of the photonic wire leads a large interaction cross-section with \emph{both} QD transitions.

The experimental set-up is depicted in Fig. \ref{fig:1}b.
It is based on a  micro-photoluminescence set-up at a temperature of T=6K.
%featuring the possibility of resonant addressing of both transitions of the X-XX three-level scheme.
%The three-level structure is obtained via the biexcitonic-excitonic (XX-X) cascade of a neutral QD. %\cite{Moreau},
By first performing standard photoluminescence (PL) measurements of an individual QD using  non-resonant laser excitation, we are able to identify the neutral X and XX transition energies, separated by $0.6$ meV around $1.36$ eV.
Using a pulsed Titanium Sapphire laser on another set-up, we have  measured lifetimes of $1.4$ ns ($0.7$ ns) for the excitonic (biexcitonic) level of the investigated QD.

As shown in Fig. \ref{fig:1}, we can perform resonant excitations
with two continuous wave (CW) external grating diode  lasers which can be finely tuned around each
 transition of the three-level scheme.
Owing to the non-perfect QD circular symmetry, the excitonic level features a fine structure splitting of $25\;\mu$eV    of the bright excitons (denoted X$_x$ and X$_y$), corresponding to two orthogonal optical dipoles oriented along the cristallographic axis $x = [110]$ and $y=[1$-$10]$ of GaAs.   In our experiments, the main qualitative effects are explained by only one (X$_y$) of the two dipoles, but the quantitative modeling requires the inclusion of both excitonic levels \cite{SM}.
The X$_y$  dipole is oriented so that   it exhibits a non-vanishing angle $\alpha=27^o$ with the laser beams' linear polarization (see Fig. \ref{fig:1}c).
Thanks to a polarizing beam splitter, laser parasitic reflections are suppressed by a factor of $10^{-4}$ \cite{Kuhlmann13}, while the light emitted by the QD dipoles on the orthogonal polarization is detected   by a charge coupled device (CCD) at the output of a grating spectrometer featuring a $12\;\mu$eV  spectral resolution.
The cross non-linear effect is revealed by measuring the reflectivity of one of the laser beams (probe beam) as a function of the intensity of the other one (control beam). We discuss below the two scenarios corresponding to the control  laser being tuned either around the upper or the lower   transition of the three-level scheme.

\begin{figure*}%[t]
\center{\includegraphics[width=0.7\linewidth]{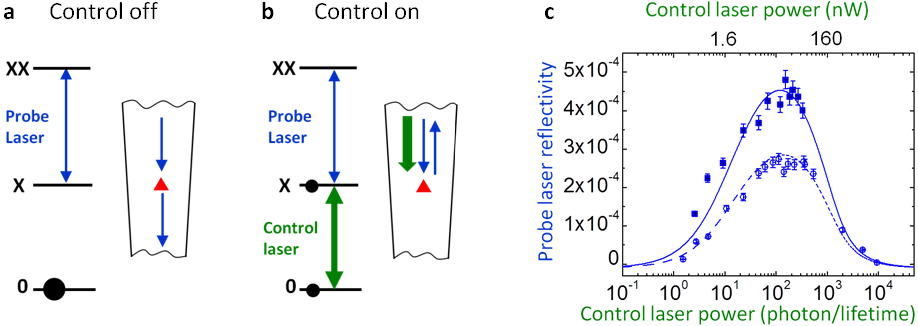}}
\caption{\textbf{Population switch.}  The control (probe) laser is tuned around the lower (upper) transition.
\textbf{a}, When the control laser is off, the probe laser beam sees an empty transition and is transmitted.
 \textbf{b}, When the control laser is on, the excitonic state is populated so that the probe laser beam is  reflected.
 \textbf{c},  Probe reflectivity as a function of the control laser power,
for a probe laser power of  $0.5$ nW (solid squares) and of $2.6$ nW (empty circles). This corresponds respectively to $0.1$ and $0.5$ of the saturation power. The solid and dashed lines are fits using our theoretical model with parameters that have also been used for the fitting of the data presented in Fig.~\ref{fig:AT}  }
\label{fig:reversed}
% Fig.3b, probe power is 0.5 nW (P/Psat~0.1), and control power is 2.5 nW.
\end{figure*}

We first consider the case in which the control (probe) laser is tuned on the lower (upper) transition (see Fig. \ref{fig:reversed}).
We will refer to this configuration as the "population switch", since
the physics at work in this situation is  the control of the X state  population by the control laser. When the latter is off (Fig. \ref{fig:reversed}a), both X and XX states are empty so that the probe laser beam sees a transparent medium and is totally transmitted.
As the control laser intensity is increased towards saturation of the lower transition, the X state becomes populated so that the probe  beam experiences a dipole-induced reflection (Fig. \ref{fig:reversed}b).

This "population switch" mechanism is evidenced in Fig.~\ref{fig:reversed}c, which shows the probe laser reflectivity as a function of the control laser power. The switching threshold is only $1.6$ nW ($10$ photons/lifetime), which  is comparable to the best results obtained recently with a single molecule \cite{Maser16}. The probe reflectivity reaches a maximum for a  control laser power as low as $16$ nW ($100$ photons/lifetime). Increasing further the control laser power leads to an Autler-Townes splitting \cite{Autler55,Cohen92,Xu07,Imamoglu08} of the intermediate state, which brings the probe beam out of resonance and reduces its reflectivity (Fig.~\ref{fig:reversed}c).
%Compared to the Autler-Townes configuration, the switching appears  here for a much lower power of about 2 nW (about 10 photons/lifetime).

This experimental behaviour is well fitted with  our  model  over  4 orders of magnitude of control laser power, for two different probe powers (see Fig.~\ref{fig:reversed}c and Supplemental Material \cite{SM}).
The values of the reflectivity $R$ and the switching power $P_{s}$ are  presently limited by imperfections of our system, and are fully accounted for  by our model.
The parameters that are affecting the performances  are the fraction $\varepsilon$ of input light coupled to the QD, and the linewidth broadening.
The quantity $\varepsilon$ is the product of the mode matching efficiency, the taper modal efficiency and  the waveguide coupling efficiency $\beta$.
  From the global fitting of our experimental results, we extract $\varepsilon =0.26\pm0.01$, which is in line with a Fourier modal method calculation  \cite{Hayrynen16} based on the sample geometry. In our experiment, the measured linewidth $\Gamma$ is broadened to $ \Gamma=10 \gamma$, where $\gamma$ is the lifetime limited linewidth. This broadening is shared between homogeneous origin due to pure dephasing, and inhomogeneous origin caused by spectral diffusion \cite{SM}.

With ideal parameters ($\varepsilon=1$  and  $\Gamma=\gamma$), losses are vanishing, so that all input light is used to saturate the QD, and the cross polarized detection scheme can be removed by  aligning the laser polarizations to the exciton dipole direction and detect the reflected light along this polarization as well. In this case, based on our theoretical model, we find that the switching power can be as low as $P_s^{(0)}=0.1$ photons/lifetime but that the maximum reflectivity can never exceed $R^{(0)}=0.1$. This limitation comes from  the  partial population of the $X$ state at low control powers and the  Autler-Townes induced probe laser detuning for higher control powers. It is also due to the population leaks caused by the presence of the other fine structure split level  X$_x$, which is populated  via spontaneous emission from the biexcitonic (XX) state  \cite{SM}.

% and whose contribution is detected as well

% From the data fitting, we have estimated that this line broadening is distributed for one quarter into homogenous pure dephasing and for three quarters into inhomogeneous spectral diffusion.
% cite Jacek FWM paper

\begin{figure*}
\center{\includegraphics[width=0.8\linewidth]{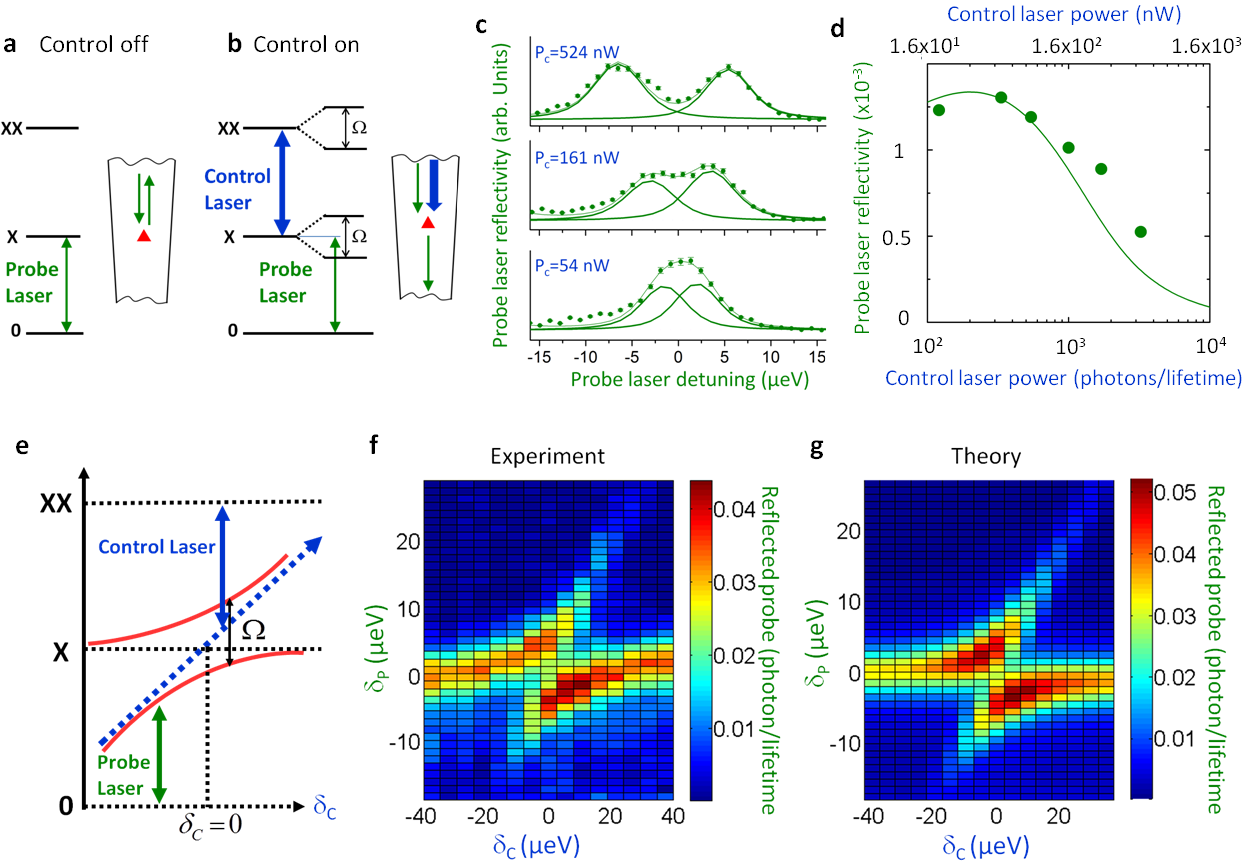}}
\caption{ \textbf{Switch in the Autler-Townes configuration}.  The control (probe) laser is tuned on the upper (lower) transition.
\textbf{a}, when the control laser is off, the probe is reflected.
\textbf{b}, when the control laser is on, it splits the $X$ state, so that the probe is no longer on resonance, and therefore transmitted.
\textbf{c}, Probe reflectivity as a function of its detuning for different control laser powers, and a zero control laser detuning.
The solid line are  fits using individual line profile   with the line position as a free parameter. The thinner line is the sum of the  individual lines.
We have checked that the splitting scales as the square root of the control laser power (data not shown) \textbf{d}, Probe reflectivity as a function of the control laser power. The probe laser power is $1$ nW. The solid line is the result given by our theoretical model.
  \textbf{e}, Position of the Autler-Townes doublet as the control laser is scanned across the upper transition. \textbf{f}, experimental \cite{SM}, and \textbf{g}, theoretical reflectivity  of the probe laser beam as a function of probe and control laser detunings. The probe (control) laser power is $1$ nW ($274$ nW).
  The theoretical fits are all made with the same set of parameters as in Fig.\ref{fig:reversed}c.}
\label{fig:AT}
\end{figure*}

To overcome this fundamental limitation, we explore another switch mechanism  in which the control (probe) laser is tuned on the upper (lower) transition (see Fig. \ref{fig:AT}). The physical effect  is here the dressing of the upper transition by the control laser when it is well above saturation.
In  the absence of the control laser (Fig. \ref{fig:AT}a),  the weak probe beam (below saturation) is reflected when on resonance with the lower transition \cite{Alexia07}.
When the control laser is turned on above saturation (Fig. \ref{fig:AT}b,c), the X$_y$ state  splits into two dressed states, as a result of the Autler-Townes effect \cite{Autler55,Cohen92,Xu07,Imamoglu08}. The probe beam is then no longer resonant and
its reflection  switched off.
In this "Autler-Townes" configuration, the  switching threshold is found to be around $200$ nW (cf. Fig. \ref{fig:AT}d).
The Autler-Townes splitting effect is well evidenced in Fig. \ref{fig:AT}e-g)  exhibiting the typical anticrossing of the probe laser reflectivity as a function of the two laser detunings.
 Using the same set of parameters for all the data presented in this work, our theoretical model is able  to quantitatively reproduce  all the experimental results  (Figs. \ref{fig:reversed},\ref{fig:AT}).

% fig2c 1nW ~ 6.3 photons/lifetime (P/Psat ~ 0.2), and control power is 274 nW.
%Autler-Townes splitting

Interestingly, and  contrary to the "population switch" configuration, the "Autler-Townes" configuration potentially shows perfect performances (i.e. $R^{(0)}=1$, $P_s^{(0)}=1$ photons/lifetime) with ideal parameters ($\varepsilon=1$ and $ \Gamma=\gamma$) in the copolarized setting. Moreover, our theoretical model indicates that, in this case, the  reflectivity is almost fully coherent, which is a key feature  in the perspective of the realization of quantum logical gates  \cite{SM}.

Using models developed by some of us in \cite{deSantis17}, we have also theoretically investigated the pulsed situation for both configurations. We have found that the pulsed regime leads to similar performances for reflectivity switching threshold and predict a pulse bandwidth of a few  $10$ MHz, set by the QD lifetime.

Let us mention that  combining   state of the art optical coupling with a narrow line QD  would allow us to
come rather close to these ideal parameters and to implement  efficient  ultra-low power all-optical switch.  Optical couplings as high as $\varepsilon =0.75$ have already been reported  in slightly narrower  waveguides  and expected values for optimized designs are as high as $\varepsilon =0.95$ \cite{Munsch13}. Additionally, close to lifetime-limited linewidths (i.e. $\Gamma \approx \gamma$) have been obtained
 recently  by applying a voltage bias across the QD \cite{Somaschi16,Kuhlmann15}. Suitable electrical contacts  can be implemented on our photonic wire, without degrading the optical properties, using the   designs proposed by some of us in \cite{Gregersen10}.

In conclusion, we have experimentally demonstrated a giant \emph{two-modes} cross non-linearity between two different laser beams in a semiconductor QD embedded in a photonic wire. This non-linearity appears at an optical  power as low as  $10$ photons  per emitter lifetime paving the way for the realization of classical, as well as quantum, ultra-low power logical gates.
Importantly, our results  can be readily transferred to planar GaAs photonic chips based on ridge waveguides \cite{Makhonin14,Stepanov_ridge} or  photonic crystals geometries \cite{Bose12,Javadi15},  and offer therefore interesting perspectives for on-chip photonic computation.

%The splitting is given by the Rabi frequency and proportional to the square root of the control laser intensity as shown in Fig.\ref{fig:2}e. This AT effect is illustrated in Fig.~\ref{fig:2}(b), which shows the evolutions of two dressed states, when the control laser scans across the $XX$ level.

%In this configuration, the switching effect appears for control laser power larger than $200$ nW ($3$ orders of magnitude above saturation), and moreover with an imperfect contrast.

\section*{Acknowledgment}
The authors wish to thank E. Wagner for technical support in data acquisition. Sample fabrication was carried out in the \textit{Upstream Nanofabrication Facility} (PTA) and CEA LETI MINATEC/DOPT clean rooms.
 H.A.N. was supported by a PhD scholarship from Vietnamese government, T.G. by the ANR-QDOT (ANR-16-CE09-0010-01) project, P.L.d.A. by the ANR-WIFO (ANR-11-BS10-011) project and CAPES Young Talents Fellowship Grant number 88887.059630/2014-00,  D.T. by a PhD scholarship from the Rh\^{o}ne-Alpes Region, and N.G.  by the Danish Research Council for Technology and Production (Sapere Aude contract DFF-4005-00370).

%\subsection*{}
%See Supplement 1 for supporting content.

\newpage

\part*{Supplemental material}
\setcounter{part}{0}

In this supplemental material, we give details on the sample geometry, we present the theoretical model and explain how the experimental parameters used in the model are obtained from the experimental data, we give some extra experimental details, and finally we give the performance of our system in ideal conditions.

\section{A quantum dot in a photonic wire}

The sample studied in this work is made of epitaxial GaAs and has the shape of an inverted cone lying on a pyramidal pedestal. The inverted cone is 17.2 $\mu$m  high, the diameter at the waist is 0.5 $\mu$m and the top facet diameter is 1.9 $\mu$m. This "photonic trumpet" embeds a single layer of a few tens of self-assembled InAs QDs, located 0.8 $\mu$m above the waist (position determined by cathodoluminescence).  To optimize the light extraction efficiency, the top facet is covered by an anti-reflection coating (115 nm thick Si$_{3}$N$_{4}$ layer). To suppress spurious surface effects, the wire sidewalls are passivated  with a 20 nm thick Si$_{3}$N$_{4}$ layer. We define such structures with a top-down process, very similar to the one described in \cite{Munsch13}.

\section{Theoretical model}
\label{theo}

\paragraph{Master equation for the electronic system.}
We consider the 4-level system formed by the electronic eigenstates $| 0 \rangle,|\text{X}_x\rangle, |\text{X}_y\rangle,|\text{XX}\rangle$. Under laser driving, the system Hamiltonian can be written as
\begin{equation}
H= H_0 + H_L ,
\end{equation}
where $H_0$ represents the electronic part, and $H_L$ the interaction with a coherent optical drive within the rotating-wave approximation (see e.g. \cite{cohen})
\begin{equation}
H_0 = E_{\text{X}_x} |\text{X}_x\rangle \langle \text{X}_x | + E_{\text{X}_y} |\text{X}_y\rangle \langle \text{X}_y | + E_{\text{XX}} |\text{XX}\rangle \langle \text{XX} |
\end{equation}

\begin{equation}
H_L  = \frac{\hbar\Omega_1}{2}e^{i \omega_1 t} |0\rangle \langle \text{X}_h|
+  \frac{\hbar\Omega_2}{2}e^{i \omega_2 t}
|\text{X}_h\rangle \langle \text{XX} | + \text{h.c.} ,
\label{HL}
\end{equation}
where $h$ ($v$) is the polarization along ( perpendicular to) the lasers, and $|\text{X}_{h}\rangle = \cos \alpha |\text{X}_y\rangle + \sin \alpha |\text{X}_x\rangle$  the exciton mode along this polarization ($|\text{X}_{v}\rangle = \sin \alpha |\text{X}_y\rangle - \cos \alpha |\text{X}_x\rangle$).
The quantities $\omega_{1,2}/2\pi$  are the laser frequencies, and $\Omega_{1,2}/2\pi$ the corresponding Rabi Frequencies.
The Lindblad master equation for the density matrix $\rho$ reads \cite{Carmichael}
\begin{equation}
\frac{\partial \rho}{\partial t} = \frac{i}{\hbar} [\rho,H] + \mathcal{L}_{\text{decay}} + \mathcal{L}_{\text{dephasing}} ,
\label{lindblad}
\end{equation}
where $\mathcal{L}_{\text{decay}}$ describes the radiative decay processes and $\mathcal{L}_{\text{dephasing}}$ the pure dephasing processes. They read respectively
\begin{equation}
\begin{split}
\mathcal{L}_{\text{decay}} & = \gamma L_{|\text{X}_x\rangle \langle 0 |}(\rho)
+ \gamma L_{|\text{X}_y\rangle \langle 0 |}(\rho) \\
& +\gamma L_{|\text{XX}\rangle \langle \text{X}_x |}(\rho)
+\gamma L_{|\text{XX}\rangle \langle \text{X}_y |}(\rho) \\
\end{split}
\end{equation}
\begin{equation}
\begin{split}
\mathcal{L}_{\text{dephasing}} & = \frac{\gamma^*}{2}  L_{|\text{X}_x\rangle \langle \text{X}_x |+|\text{X}_y\rangle \langle \text{X}_y | - |0\rangle \langle 0 |}(\rho) \\
& + \frac{\gamma^*}{2}  L_{|\text{XX}\rangle \langle \text{XX} |-|\text{X}_x\rangle \langle \text{X}_x |-|\text{X}_y\rangle \langle \text{X}_y |}(\rho) \\
& + \frac{\gamma^*}{2}  L_{|\text{XX}\rangle \langle \text{XX} | - |0\rangle \langle 0 | }(\rho) , \\
\end{split}
\end{equation}
where $L_C(\rho)$ is the Lindblad superoperator for a collapse operator $C$:
\begin{equation}
L_C(\rho) =  \left[C \rho C^{\dagger} - \frac{1}{2} \left(\rho C^{\dagger} C + C^{\dagger} C \rho \right) \right] .
\end{equation}
In the radiative decay term $\mathcal{L}_{\text{decay}}$, we have assumed that the four different possible exciton recombination processes occur at the same rate $\gamma$. In the pure dephasing term $\mathcal{L}_{\text{dephasing}}$, an additional decay of the coherence between the ground state, the excitonic states and the biexciton state is considered with a rate $\gamma^*/2$, corresponding to an additional spectral broadening of $\hbar \gamma^*$ for the full width at half maximum (FWHM) of these transitions.

\begin{figure}
\center{\resizebox{0.27 \textwidth}{!}{\includegraphics{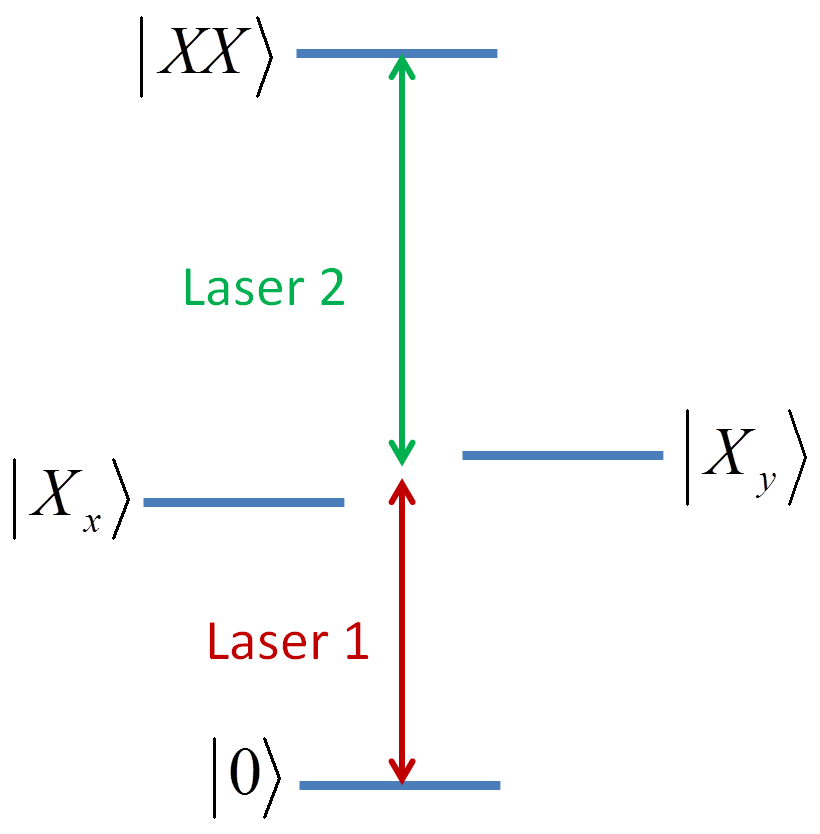}}}
\caption{ \textbf{Level scheme}.
Levels $|\text{X}_x\rangle $ and $|\text{X}_y\rangle $ are the two fine structure split excitonic eigenstates.
In practice they are separated by $25 \; \mu $eV, and we can consider only one of them.  Level $|\text{XX}\rangle $ corresponds to the biexciton. Laser 1 (2) is tuned around the lower (upper) transition.
}
\label{fig:niveaux}
\end{figure}

\paragraph{Optical nonlinearities.}
The optical nonlinearities are calculated following previous studies of quantum electrodynamics in one-dimensional wave-guides \cite{Auffeves07,valente12}.
The total reflected intensity (in photons per unit time) for the transition 1 (2) is proportional to the exciton (biexciton) population:
\begin{equation}
I^{\text{R}}_{1,p} =  \varepsilon \frac{\gamma}{2} \hspace{0.1em} n_{\text{X}_p} ,
\label{T1tot}
\end{equation}
\begin{equation}
I^{\text{R}}_{2,p} =  \varepsilon \frac{\gamma}{2} \hspace{0.1em} n_{\text{XX}} ,
\label{T2tot}
\end{equation}
where $p=v$ or $h$ is the polarization,  $n_i=\rho_{ii}$ is the population of the level $i$,
and $\varepsilon$ is the
product of the mode matching efficiency with the waveguide
coupling efficiency $\beta$ (see main text).
The coherent part of these reflected intensities reads
\begin{equation}
I^{\text{R}\text{coh}}_{1,p} =  \varepsilon \frac{\gamma}{2} \hspace{0.1em} |\langle 0 | \rho | \text{X}_p \rangle|^2 ,
\label{T1coh}
\end{equation}
\begin{equation}
I^{\text{R}\text{coh}}_{2,p} = \varepsilon \frac{\gamma}{2} \hspace{0.1em} |\langle \text{X}_p | \rho | \text{XX} \rangle|^2 .
\label{T2coh}
\end{equation}
In the polarization $v$ (orthogonal to the excitation's one), the transmitted intensity is equal to the reflected one $I^{\text{T}}_{v} = I^{\text{R}}_{v}$ since the quantum dot (QD) luminescence is distributed symmetrically between the two directions without any interference with the incident beam.
In contrast, for the polarization $h$ (parallel to the excitation's one), an interference effect occurs between the incident field and the scattered field. The coherent and total (i.e. coherent + incoherent) transmitted electric fields are given respectively for the transitions 1 and 2 by
\begin{equation}
I^{\text{T}\text{coh}}_{1,h} = \varepsilon \Big\vert \sqrt{\varepsilon I_1^{\text{in}}} + i \sqrt{\frac{\gamma}{2}} \langle 0 | \rho | \text{X}_h \rangle \Big\vert^2
\end{equation}

\begin{equation}
I^{\text{T}}_{1,h} =  \varepsilon \left[ I_1^{\text{in}} + \frac{\gamma}{2} n_{\text{X}_p} + \sqrt{2\gamma I_{\text{in}}} \;\text{Im}\left( \langle 0 | \rho | \text{X}_h \rangle \right) \right]
\end{equation}

\begin{equation}
I^{\text{T}\text{coh}}_{2,h} = \varepsilon \Big\vert \sqrt{\varepsilon I_2^{\text{in}}} + i \sqrt{\frac{\gamma}{2}} \langle \text{X}_h | \rho | \text{XX} \rangle \Big\vert^2
\end{equation}

\begin{equation}
I^{\text{T}}_{2,h} =  \varepsilon \left[ I_2^{\text{in}} + \frac{\gamma}{2} n_{\text{XX}} + \sqrt{2\gamma I_2^{\text{in}}}\; \text{Im}\left(\langle \text{X}_h | \rho | \text{XX} \rangle \right) \right]
\end{equation}

\paragraph{Spectral diffusion}
In addition to pure dephasing, we consider spectral diffusion processes, i.e. fluctuation of the electronic transition energies over timescales which are long compared to the exciton lifetime $1/\gamma$. A fluctuating term $\delta E_{\text{X}} $ in the exciton energy is added as:
\begin{equation}
E_{\text{X}_i} = E_{\text{X}_{i}}^0 + \delta E_{\text{X}}
\end{equation}
\begin{equation}
E_{\text{XX}} = E_{\text{XX}}^0 + 2 \delta E_{\text{X}}
\end{equation}
where $i=x,y$, and $E_{\text{X}_{i}}^0 $ ($E_{\text{XX}}^0$) is the mean exciton (biexciton) energy.
This fluctuation is assumed to verify a Gaussian distribution, with a probability density function:
\begin{equation}
\mathcal{P}(\delta E_{\text{X}}) = \frac{1}{\sigma_{\text{X}} \sqrt{2\pi}} \exp \left[-\frac{\delta E_{\text{X}}^2}{2\sigma_{\text{X}} ^2} \right]
\end{equation}
where $\sigma_{\text{X}}$ is the standard deviation. The corresponding FWHM reads $w_{\text{diff}} = 2 \sqrt{2 \ln (2)} \sigma_{\text{X}}$. The total FWHM of the excitonic and biexcitonic transitions reads
\begin{equation}
\Gamma = \gamma + \gamma^* + w_{\text{diff}}
\end{equation}

For each realization of spectral diffusion, we calculate the density matrix $\rho(\delta E_{\text{X}})$ under continuous-wave excitation by finding the steady-state of the above Lindblad equation (Eq.~\ref{lindblad}). The density matrix and the reflectivities are then averaged %over the sampling
over the distribution $\mathcal{P}(\delta E_{\text{X}})$.

\paragraph{Effective 3-level system.}
For Rabi frequencies and detunings which remains small compared to the fine structure splitting ($E_{\text{X}_y}-E_{\text{X}_x}=27\;\mu$eV), the system behaves as a 3-level system ($| 0 \rangle, |\text{X}_y\rangle,|\text{XX}\rangle$), i.e. the influence of the state $|\text{X}_x\rangle$ is negligible.
The light-matter interaction can be written
\begin{equation}
H^{\text{eff}}_L  = \frac{\hbar\Omega^{\text{eff}}_1}{2}e^{i \omega_1 t} |0\rangle \langle \text{X}_y|
+  \frac{\hbar\Omega^{\text{eff}}_2}{2}e^{i \omega_2 t}
|\text{X}_y\rangle \langle \text{XX} | + \text{h.c.} ,
\label{HL_eff}
\end{equation}
with $ \Omega^{\text{eff}}_i =  \Omega_i \sin \alpha $.
In addition, a factor  $\cos \alpha$ (or $\cos^2 \alpha$) appears when expressing the coherences (or populations) of the horizontal exciton mode for cross-polarized reflections, such as
$\langle 0 | \rho | \text{X}_h \rangle =  \langle 0 | \rho | \text{X}_y \rangle \cos \alpha$ and $n_{\text{X}_h} =  n_{\text{X}_y}\cos^2 \alpha$ .

\section{Extraction of experimental parameters}

The unknown parameters of the QD-waveguide system are the input-output coupling efficiency $\varepsilon$, the pure dephasing rate $\gamma^{*}$ and the spectral diffusion factor  $w_{\text{diff}} $. The factor $\varepsilon$ is extracted very precisely from the Autler-Townes splitting results (see  Fig. 3 of the main text), wherein the control laser is well above saturation and dresses the upper transition for one of the fine structure split level ($|\text{X}_y\rangle$ for example), while the well below saturation probe laser probes the Autler-Townes splitting induced by the control laser. The Autler-Townes splitting is given by $\Omega=\sqrt{\delta^{c}_{2}+\Omega^{2}_{c}}$, where $\delta_c$ is the detuning between the control laser  and the $|\text{X}_y\rangle - |\text{XX}\rangle$ transition, and $\Omega_{c}$ is the Rabi frequency of  the control laser that directly interacts with the QDs inside the photonic wire. We have $\Omega_{c}^2=2 \varepsilon \gamma I^{\text{in}}_c$, where $I_c^{\text{in}}$ is the control laser  intensity (in photons per excitonic lifetime) at the input of the photonic wire \cite{valente12}. By measuring $\Omega$ at zero-detuning  ($\delta_{c}=0$), and comparing with $I_c^{\text{in}}$,  we find a coupling efficiency factor $\varepsilon =0.26\pm0.01$.

The next unknown factors are the pure dephasing $\gamma^{*}$ and the broadening induced by spectral diffusion $w_{\text{diff}} $ (see section \ref{theo} above). A total linewidth of  $4 \mu$eV has been measured from the resonant laser scans (see Fig. 3c,f of the main text).
The weight between the two broadening mechanisms is then determined by optimizing the fits with the experimental results. This leads us to take $\gamma^{*}=1\pm0.5\mu$eV and $w_{\text{diff}} =3\pm0.5\mu$eV, where the lifetime limited linewidth is $\gamma=0.5\mu$eV.

\section{Measurement of the probe reflectivity}
%\label{AT}

For a given set of control and probe powers, the best probe reflectivity is measured at the best laser detunings from the two-dimensional reflectivity plots such as  shown in Fig.3f of the main text and Fig.\ref{fig:S2} of this supplementary material.
For these plots, the two laser frequencies are respectively scanned across the frequencies of lower and upper transitions while the probe reflectivity  is recorded.
At each step of the scan, the spectrum of the light reflected from the trumpet is recorded on a CCD camera with a typical integration time of $0.1$s. The probe reflectivity is obtained by integrating the  counts within a fixed spectral interval of more than $100 \mu$eV containing the probe laser scan interval (max $80 \mu$eV), and including therefore both fine structure split levels (splitting around 25$\mu$eV). In some cases, mainly in the population switch configuration,  both excitonic states are populated via the biexcitonic state, so that incoherent photons involving the non-principal excitonic state are also detected. Note that this feature is included in our theoretical model.

\section{Importance of the non-resonant laser}
The obtention of narrow resonant lines of the QD transitions requires
 the presence of a non-resonant laser ($\lambda=825$nm, i.e. $E=1.5$eV). The power of this laser is always kept around $0.1$nW (i.e. $10^{-3}$ times the saturation power), so that its induced photoluminescence remains negligible compared to the resonant laser luminescence.
 The carriers generated in the wetting layer by this weak non-resonant laser have been shown to reduce spectral diffusion  and hence reduce  the linewidth \cite{Majumdar11}. This is a standard technique in resonant excitation experiments with individual QDs.
 In our case the linewidth is reduced by a factor of $3$, from $15\mu$eV down to $5 \mu$eV.

\section{Experimental data including the fine structure splitting}

\begin{figure}%[t]
\includegraphics[width=\linewidth]{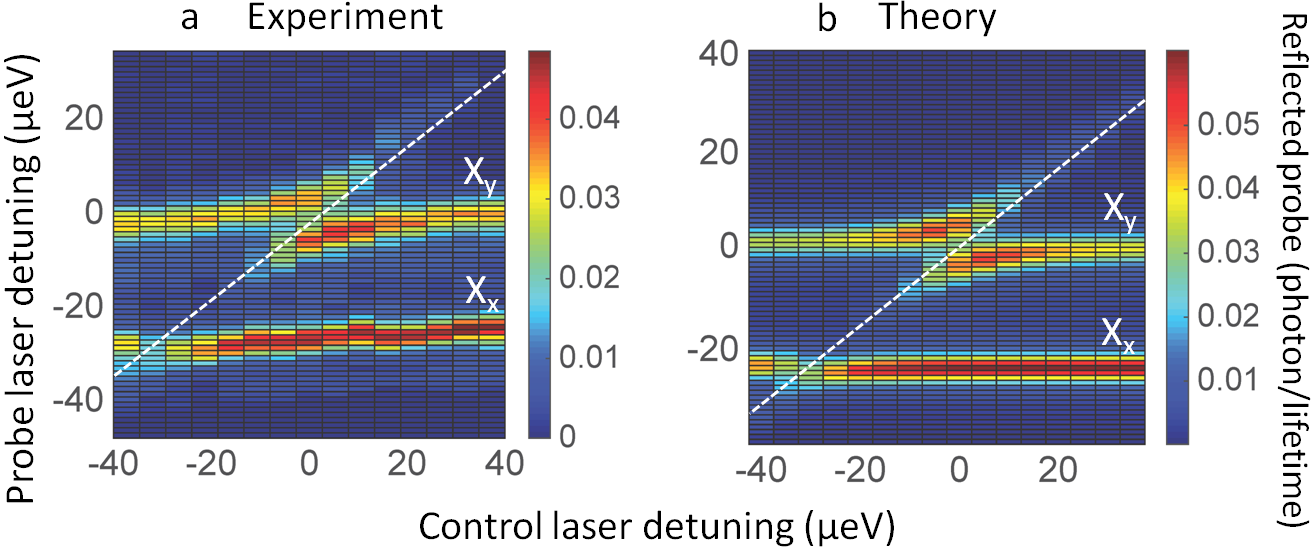}
\caption{\textbf{Fine structure splitting.} Experimental (left) and theoretical (right) results following Autler-Townes splitting approach including both excitonic levels X$_x$ and X$_y$ split by $25\;\mu$eV. The probe reflected intensity is plotted as function of control and probe detunings. The control laser power is set at $274$ nW. The probe laser scanning range is 80 $\mu$eV. The white dashed lines are guides to the eye  corresponding to the positions of the control laser beam during the scan. The model uses the coupling efficiencies $\varepsilon_{in}=\varepsilon_{out}=0.26$ and the total linewidth broadening $4 \;\mu$eV, including $1\; \mu$eV pure dephasing and 3 $\mu$eV spectral diffusion, as for all the data presented in this work.}
\label{fig:S2}
\end{figure}
In this section, we give a complete picture concerning the presence of the other excitonic dipole in the Autler-Townes splitting approach.
 To this end, we have carried out a broad scan of the probe beam covering both  excitonic states $X_{x}$ and $X_{y}$. Fig.~\ref{fig:S2} shows the experimental and theoretical results for a scan with a control laser power $274$ nW. The Autler-Townes splitting is  observed for both states. It should be mentioned  that the Autler-Townes splittings are different for each state,  since, except for $\alpha = 45^{o}$, the two orthogonally polarized excitonic dipoles are excited differently with respect to the resonant laser polarization. The Autler-Townes splittings at resonance for the two dipoles, $\Omega_{x}$ and  $\Omega_{y}$ scale as  $\Omega_{y}/\Omega_{x}=1/\tan  \alpha$, with $\alpha$ the angle between the polarizations of laser and $X_{y}$ as defined in Fig. 1 of the main text. In our experiment, we have chosen  $\alpha=27^{o}$ so that   $\Omega_{y}/\Omega_{x}\approx2$.

\section{Switch performances in ideal conditions}

In this section we compute the reflectivity with ideal parameters ($\varepsilon=1$ and $ \Gamma = \gamma $). The total reflectivity is given by equations (\ref{T1tot},\ref{T2tot}), whereas the coherent part is given by equations (\ref{T1coh},\ref{T2coh}).

\subsection{Population switch configuration}

\begin{figure}%[t]
\resizebox{0.53 \textwidth}{!}{\includegraphics{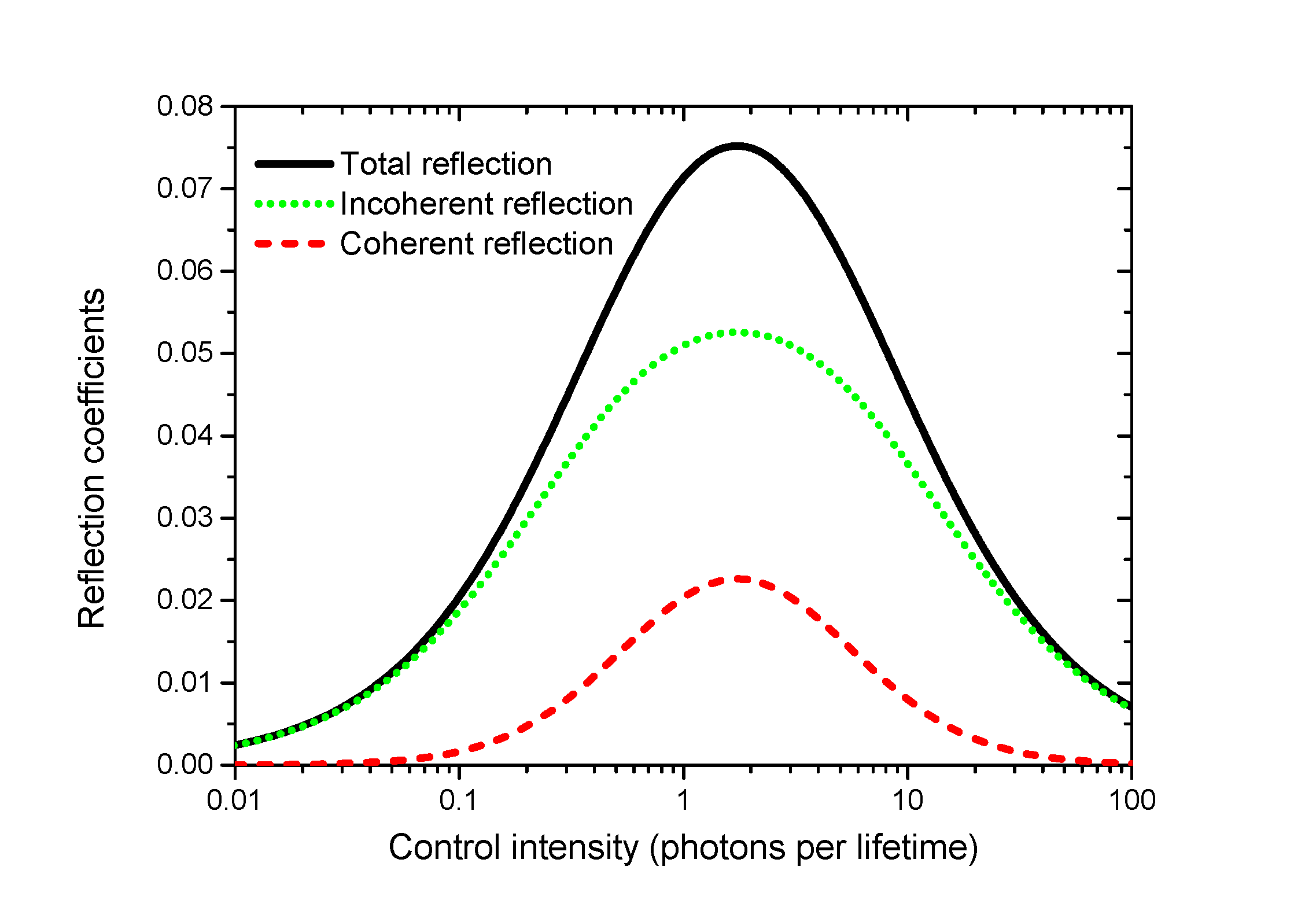}}
\caption{\textbf{Ideal probe reflectivity in the population switch case}. The computed reflectivity of a vanishingly weak probe laser (tuned on the upper transition) is plotted as a function of the control laser power (tuned on the lower transition)  in  a co-polarized situation with ideal parameters ($\varepsilon=\Gamma/\gamma=1$). The probe power is $10^{-3}$ photon/lifetime. It can be observed that the reflection is mainly incoherent.
}
\label{ideal_PW}
\end{figure}

In this approach, the control (probe) beam is tuned on the lower (upper) transition (cf. Fig. 2 of the main text). The switching effect is based on a population effect, so that incoherent scattering is dominant even
with ideal parameters  (see Fig. \ref{ideal_PW}). Additionally, owing to  the never complete population of the X$_y$ state at low power, the  Autler-Townes induced probe laser detuning for higher control powers, and population leaks to the other excitonic state X$_x$ via spontaneous emission from the biexcitonic state XX, the maximum reflectivity in the population switch configuration never exceeds $R=0.08$.

\paragraph{Cross-polarized.}
Here the detected light is linearly polarized orthogonally to the input laser, as in our experiment, to ensure a low level of parasitic backscattered laser light.
Note that the other fine-structure split level X$_x$, which is  also populated via the XX state, is also detected by our experimental set-up via the XX-X$_x$ spontaneous emission, and contributes therefore to the probe reflectivity in an incoherent way, so that the total probe reflectivity is mainly incoherent. For $\alpha=45^{o}$, its maximum value is found to be $R=0.03$ obtained for $1$ photon/lifetime of control laser intensity.

\paragraph{Co-polarized.}
In the case of low backscattered light from the trumpet top facet, we can in principle access the light reflected from the trumpet along the same polarization as the input laser. In this situation, the relevant QD dipole X$_y$ orientation is chosen along this polarization.
As mentioned above, even in this case, the population switch configuration does not allow for unity reflectivity (see Fig.\ref{ideal_PW}).
Moreover, since half of the reflectivity is due to the other excitonic state (X$_x$ state), via the XX-X$_x$ spontaneous emission, not more than half of the reflected light is coherent.

\subsubsection{Autler-Townes configuration}
In the Autler-Townes approach, the control (probe) is tuned on the upper (lower) transition.
As it is shown below, the Autler-Townes configuration offers very good switching performance, including coherence, for ideal parameters. The presence of the other fine structure split excitonic level has almost no effect on  performance, owing to the always low population of the biexcitonic state.

\paragraph{Cross-polarized.}
For $\alpha=45^{o}$, the maximum reflectivity is limited by the term
$ (1/2)\cos^2 \alpha \sin^2 \alpha =0.125$, where $\alpha $ is the angle of the relevant dipole with the exciting laser polarization (see Fig. 1 of the main text), and
where the $1/2$ factor comes from the fact that, owing to the cross-polarization scheme, the reflected light does not interfere with the incoming laser, so that half of the emitted light is directed towards the substrate.
 Having  the reflectivity switched to half of this ideal value requires, from our model, a control laser power of about $3$ photons/lifetime, and the reflectivity is found mainly coherent.

%\begin{figure}%[t]
%\includegraphics[width=\linewidth]{FigS3}
%\caption{\textbf{Probe reflectivity for optimized parameters of the system in crosspolarized configuration}. The probe power is set at $P_{\mbox{Probe}}/P_{\mbox{sat}} = 0.01$. The black curve plots the fitted reflectivity for the experimental condition (Fig. \ref{fig:2}), in which $\varepsilon =0.26$, and $\alpha = 27^{\degree}$, pure dephasing rate ($\gamma ^* = 1 \mu$eV , spectral diffusion width of $w_{\text{diff}}=3 \mu$eV. The red curve plots the calculated reflectivity for the same spectral broadening as the black one ($4 \mu$eV in total), but with optimum coupling rates $\varepsilon =1$. The blue curve shows the ideal cross-polarized probe reflectivity at $\alpha = 27^{\degree}$ for a perfect system with zero broadening and $\varepsilon =1$. The pink curve shows also the ideal cross-polarized probe reflectivity for a perfect system with zero broadening and $\varepsilon =1$, but with an optimum $\alpha = 45^{\degree}$.}
%\label{fig:S3}
%\end{figure}

\begin{figure}%[t]
\resizebox{0.53 \textwidth}{!}{\includegraphics{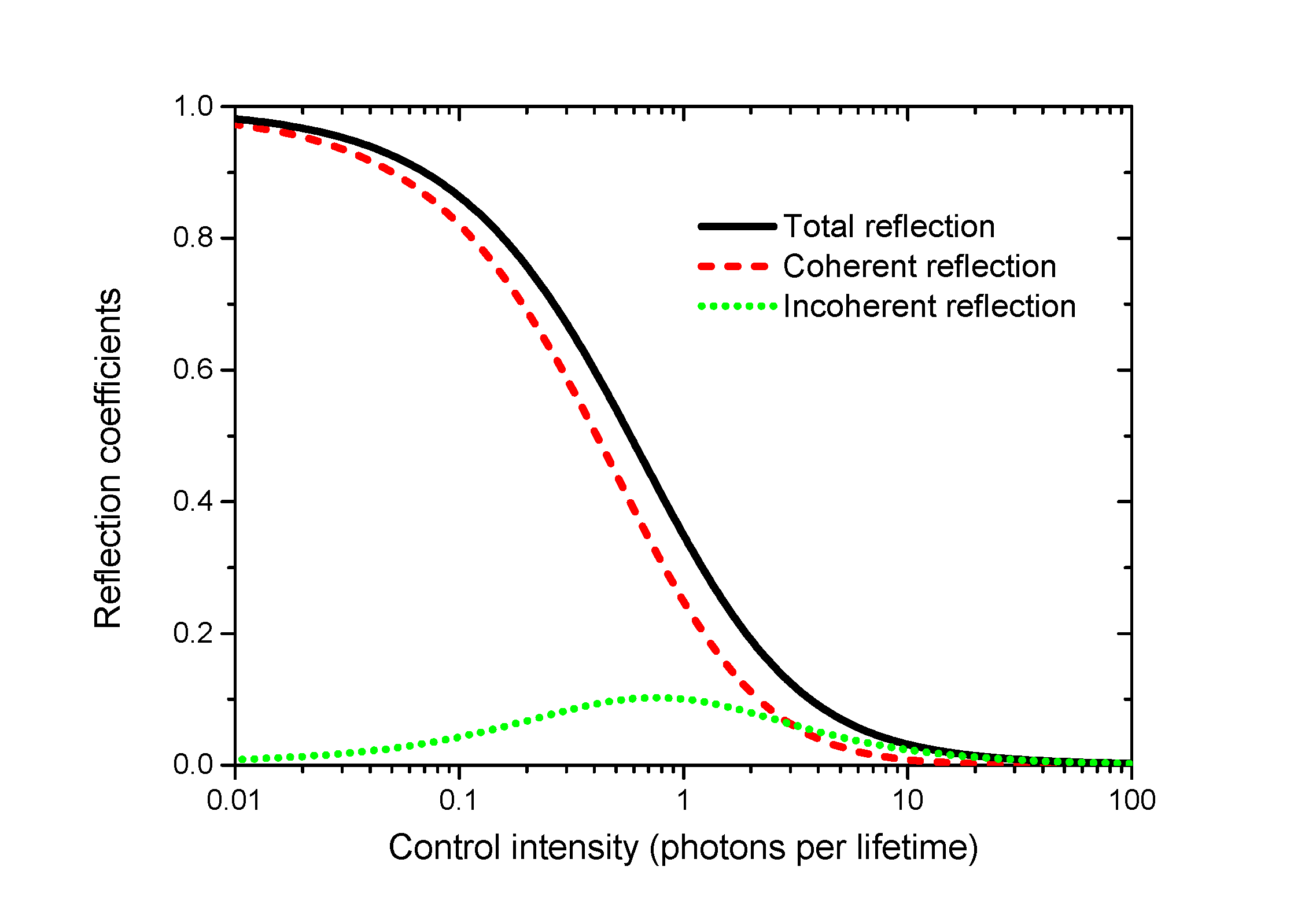}}
\caption{\textbf{Ideal probe reflectivity in the Autler-Townes configuration}.
The computed reflectivity of a vanishingly weak probe laser (tuned on the lower transition) is plotted as a function of the control laser power (tuned on the upper transition) in  a copolarized situation with ideal parameters ($\varepsilon=\Gamma/\gamma=1$). The probe power is $10^{-3}$ photon/lifetime.
The reflectivity is mainly coherent and reaches  unity at low control laser power.}
\label{ideal_AT}
\end{figure}

\paragraph{Co-polarized.}
In this configuration, the probe reflectivity reaches unity for a vanishing control laser power, and is fully coherent (see Fig.\ref{ideal_AT}). This is a key assets
for applications in quantum information processing.
The control laser power required to switch the reflectivity down to a value of $0.5$  is  $1$ photon/lifetime.

\end{document}